# Intrinsically Activated SrTiO$_3$: Photocatalytic H$_2$ Evolution from Neutral Aqueous Methanol Solution in the Absence of Any Noble Metal Cocatalyst

Xuemei Zhou,[†] Ning Liu,[†] Tadahiro Yokosawa,[‡] Andres Osvet,[§] Matthias E. Miehlich,[∥] Karsten Meyer,[∥] Erdmann Spiecker,[‡] and Patrik Schmuki*,[†,⊥]

[†]Department of Materials Science, WW4-LKO, and [§]Department of Materials Science, WW6, iMEET, University of Erlangen-Nuremberg, Martensstrasse 7, 91058, Erlangen, Germany

[‡]Center for Nanoanalysis and Electron Microscopy (CENEM), University of Erlangen-Nuremberg, Cauerstrasse 6, 91058 Erlangen, Germany

[∥]Department of Chemistry and Pharmacy, Inorganic & General Chemistry, University of Erlangen-Nuremberg, Egerlandstrasse 1, 91058 Erlangen, Germany

[⊥]Department of Chemistry, Faculty of Science, King Abdulaziz University, P.O. Box 80203, Jeddah 21569, Saudi Arabia.

*E-mail: schmuki@ww.uni-erlangen.de. Phone: +49 91318517575. Fax: +49 91318527582.

**ABSTRACT:** Noble metal cocatalysts are conventionally a crucial factor in oxide- semiconductor-based photocatalytic hydrogen generation. In the present work, we show that optimized high-temperature hydrogenation of commercially available strontium titanate ($SrTiO_3$) powder can be used to engineer an intrinsic cocatalytic shell around nanoparticles that can create a photocatalyst that is highly effective without the use of any additional cocatalyst for hydrogen generation from neutral aqueous methanol solutions. This intrinsic activation effect can also be observed for $SrTiO_3$[100] single crystal as well as Nb-doped $SrTiO_3$[100] single crystal. For all types of $SrTiO_3$ samples (nanopowders and either of the single crystals), hydrogenation under optimum conditions leads to a surface-hydroxylated layer together with lattice defects visible by transmission electron microscopy, electron paramagnetic resonance (EPR), and photoluminescence (PL). Active samples provide states in a defective matrix−this is in contrast to the inactive defects formed in other reductive atmospheres. In aqueous media, active $SrTiO_3$ samples show a significant negative shift of the flatband potential (in photoelectrochemical as well as in capacitance data) and a lower charge-transfer resistance for photoexcited electrons. We therefore ascribe the remarkable cocatalyst-free activation of the material to a synergy between thermodynamics (altered interface energetics induced by hydroxylation) and kinetics (charge transfer mediation by suitable $Ti^{3+}$ states).

KEYWORDS: *SrTiO3, hydrogenation, noble-metal-free, water splitting, intrinsic activation*

**INTRODUCTION**

Since the early 1970s, $SrTiO_3$ (STO) has been one of the most widely studied oxide semiconductors to produce photocatalytic $H_2$ from aqueous solutions.[1−4] STO has a conduction band that lies energetically negative to the $H_2/H_2O$ redox couple and thus provides the fundamental energetic prerequisite for a reaction of photogenerated electrons with water to produce hydrogen. Because of these early findings, a large number of other and often more effective semiconductors (visible light absorption, charge carrier transfer to electrolyte) have been explored;[5−9] however, STO remains one of the most investigated materials because of its favorable energetics, low cost, and a high resistance against photocorrosion and chemical corrosion.[10] Moreover, STO single crystals (STOSCs) can easily be grown and doped in a highly defined manner and are widely available, thus, extensive surface science and photoelectrochemical data on single-crystalline STO is available. In the pioneering work of Wrighton et al.[3−8] on photoelectrochemistry of STOSCs, the material showed a remarkably high photoconversion efficiency when used as a photoanode under UV illumination. Kolesar et al.[11] showed for STO anodes at zero bias voltage an external quantum efficiency of 10% (using UV illumination of 3.8 eV), which is approximately an order of magnitude higher than efficiencies reported for TiO$_2$ anodes.[12]

In photocatalytic approaches - this is without applied external electrochemical bias - the hydrogen evolution reaction in aqueous environments (with or without sacrificial agents)[13,14] is severely kinetically hampered. Therefore, in virtually all work, for example, using semiconductor particle- suspensions, cocatalysts are required to reach reasonable H$_2$ generation rates. Most commonly, these cocatalysts are costly noble metals,[15–19] such as Pt or Rh that are decorated on the semiconductor.[20,21] The key role of the cocatalyst is to mediate the transfer of photoinduced charge carriers to the electrolyte.[22,23] Over the years, cocatalyst systems have become increasingly refined. For example, work by Domen et al.[24,25] found La and Rh codoped STO to show a strongly enhanced hydrogen evolution rate. Moreover, work by Kudo et al.[16] shows that Rh (1%)-doped STO photocatalyst loaded with a Pt cocatalyst (0.1 wt %) can provide a quantum yield of 5.2% for the H$_2$ evolution reaction. However, the use of precious metals provides a considerable economic burden on photo- catalytic H$_2$ generation.

In contrast to these noble metal approaches, already very early work by Wagner and Somorjai in 1980[26,27] reported that the photocatalytic hydrogen production in the absence of a Pt cocatalyst is essentially possible under extreme conditions (water vapor and saturated NaOH). This approach was however hardly followed up because of the severe experimental conditions required.

For titania, it has been recently reported that an intrinsic defect-engineering approach can lead to a noble-metal-free photocatalytic hydrogen generation. In the case of anatase TiO$_2$ nanotubes and powders, Liu et al.[28–31] demonstrated that hydrogenated titania, the so-called "gray" titania, can show intrinsic photocatalytic H$_2$ evolution. That is, this form of mildly reduced gray titania shows intrinsic catalytic activation - this in contrast to the widely studied "black" titania introduced by Chen and Mao (obtained under strong reduction conditions)[32] that still requires a noble metal cocatalyst to show photocatalytic H$_2$ activity. For STO, some work reports the use of similarly severe reduction treatments on STO that resulted in black STO the material shows visible light absorption but all reports (see Table S1) use additional Pt loading (typically 1.0 wt %) to achieve H$_2$ evolution.

In the present work, we investigate systematically hydro- genation treatments of STO to explore the feasibility to create noble-metal-free, intrinsic activity for photocatalytic H$_2$ evolution. For this, we use not only commercially available nanosized STO powder (STONP) but also single-crystal STO[100] (STOSC) and Nb:STO[100] substrate (NbSTOSC) - the latter to investigate and extract critical factors underlying the intrinsic photocatalytic reactivity for H$_2$ evolution on defined and analytically easily accessible surfaces and to provide defined electrodes in electrochemical investigations.

**RESULTS AND DISCUSSION**

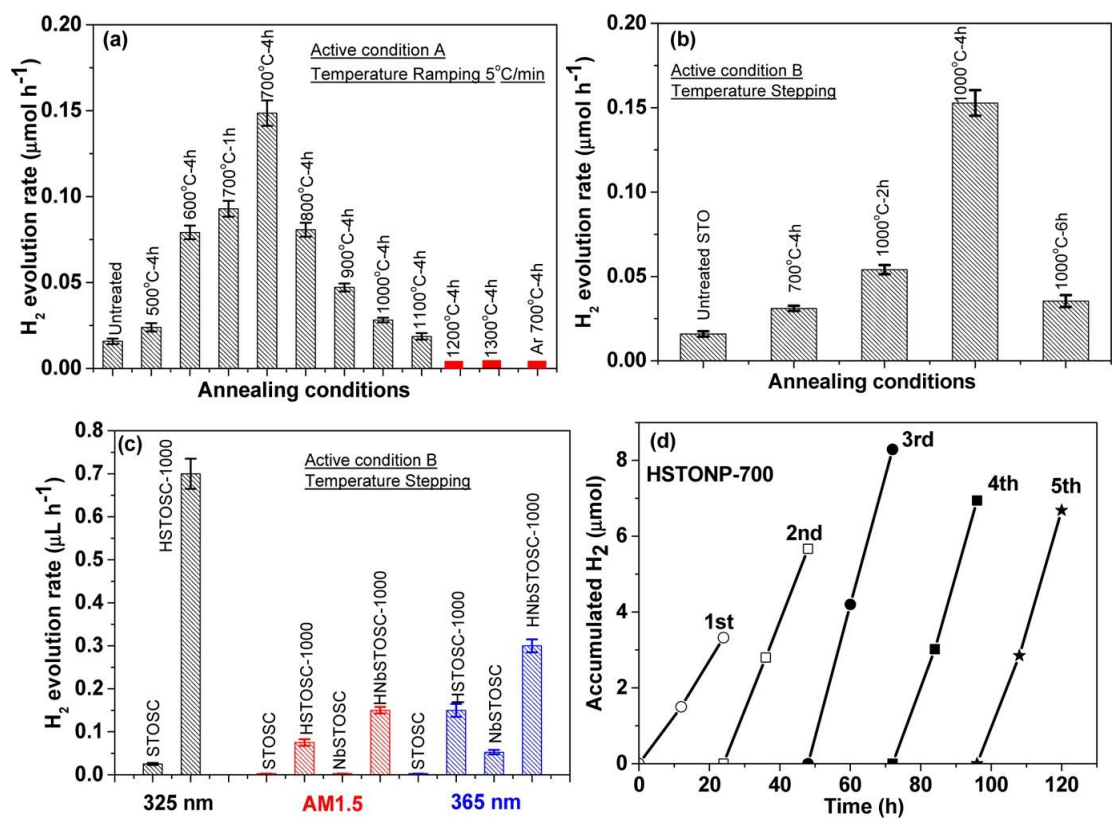

Figure 1. (a−c) Photocatalytic $H_2$ evolution from STO nanopowders and single crystals before and after annealing treatments. Red bar: no hydrogen can be detected. (a) STO nanopowders annealed in hydrogen gas with a temperature ramping of 5 °C/min (active condition A). (b) STO nanopowders annealed with temperature stepping in hydrogen (active condition B). (c) STOSCs and Nb-doped STOSCs before and after annealing in hydrogen gas at 1000 °C for 4 h with temperature stepping (active condition B). (d) Cycling $H_2$ evolution for the hydrogenated STO nanopowders at 700 °C for 4 h.

Figure 1 shows an overview of the photocatalytic hydrogen evolution rates observed after different hydrogenation treat- ments of STO powders and single crystals in a tube furnace with a $H_2$ current of ca. 6 L/h in a temperature range from 500 to 1300 °C for durations between 1 and 4 h. The nanoparticles were denoted as XSTONP-$T$, and the single crystals were denoted as X(Nb)STOSC-$T$ (as described in the Materials and Methods): where X is the gas atmosphere, H indicates hydrogen, Ar indicates argon, and $T$ is the annealing temperature. To evaluate the noble-metal-free photocatalytic performance of STO powders and single crystals, the samples were immersed/dispersed in a $H_2O$/methanol solution and illuminated under AM 1.5 (100 mW/cm$^2$) conditions or alternatively by a defined UV light source (HeCd laser or LED), and the amount of evolved $H_2$ was determined by gas chromatography (as described in the Materials and Methods)

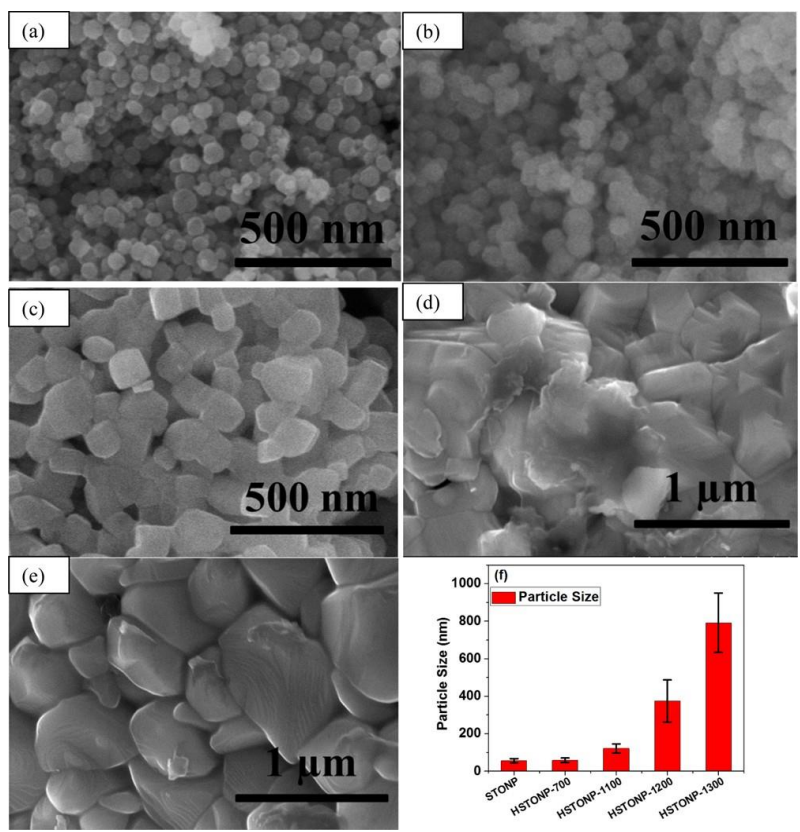

Figure 2. (a) SEM images for STO nanopowders, STO nanopowders annealed at (b) 700 °C for 4 h, (c) 1100 °C for 4 h, (d) 1200 °C for 4 h, and (e) 1200 °C for 4 h with a temperature ramping. (f) Size distribution of particles obtained from SEM images.

Evidently, hydrogen annealing can lead to activation of the powder samples (Figure 1a,b) and of both types of single- crystal samples (intrinsic STO[100], as well as Nb-doped material Nb:SrTO$_3$[100]) (Figure 1c). This activation for photocatalytic hydrogen evolution is to a large extent determined by the annealing time and temperature. For various parameter combinations, an optimum performance can be reached. For example, for the powder samples [scanning electron microscopy (SEM) images shown in Figure 2], a highest efficiency for hydrogen production is obtained by annealing the powders at 700 °C for 4 h. For higher temperature hydrogenation or longer exposure, the photo- catalytic activity of the samples drops again until virtually no activity is observed. The samples shown in Figure 1a were treated using a temperature ramping profile with holding at 700 °C (as described in the Materials and Methods). In the case of Figure 1b, a different annealing profile (direct temperature stepping) was used; with this approach, optimized conditions are obtained by annealing the sample at 1000 °C for 4 h.

We also explored if thermal reduction in H$_2$-free reductive conditions (i.e., under Ar annealing) can cause this activation effect, too. For H$_2$ and Ar heat treatments, the formation of Ti$^{3+}$/O$_v$ lattice defects because of oxygen loss from the stoichiometric compound is a common observation.[33−37] However, in our experiments by annealing in Ar, although defects were formed [as shown later in electron paramagnetic resonance (EPR) measurements], no activation for photo- catalytic H$_2$ generation could be induced (Figures 1a and S1a) in contrary, a deactivation in comparison with the untreated material was observed. Figure 1d shows a cycled measurement of the $H_2$ evolution rate, demonstrating a steady $H_2$ evolution over extended times (further long-term experimental details are given in the Supporting Information, Figure S1b,f).

The morphology of the powders for the various annealing temperatures is illustrated in the SEM images shown in Figure 2. Up to an annealing temperature of 700 °C, there is no significant change of the particle size (Figure 2b), however, at a temperature of 1100 °C or higher, the particles start sintering together and aggregate to significantly larger sizes (Figure 2). This is in line with Brunauer−Emmett−Teller (BET) measurements that show a specific surface area of 37.1 $m^2/g$ for the nontreated STO nanopowder. After hydrogenation at 700 °C (HSTONP-700), the specific area decreases to 15.1 $m^2/g$, and for 1100 °C, the surface area is below 1 $m^2/g$. Figure 1c shows that also on both type of STOSCs, hydrogen annealing can induce activation of the surface for cocatalyst-free photocatalytic $H_2$ generation. Because of the significantly lower specific surface area, the $H_2$ evolution from single crystals is correspondingly lower than from the nanoparticle suspension but the overall observations with respect to $H_2$ annealing temperature and time are in accordance with the results obtained from the powders. A discussion of the observed solar to fuel efficiencies for single crystals and powders is given in Supporting Information Figure S2.

It is noteworthy that a change of color can be seen for nanopowders (Figure S3) as well as for the single crystals (Figure S4). This can be ascribed, as widely reported in literature, to $Ti^{3+}/O_v$ that constitute color-active centers in STO.[38−44] In line, an increased visible light absorption can be observed from reflection spectra for powder (Figure S3) as well as for the nondoped STOSC (Figure S4). The Nb-doped material is already in its nontreated state strongly absorbing in the visible range because of the high doping concentration and the accordingly high electron density in the conduction band at room temperature (RT). However, plain visible light excitation was not found to be active for $H_2$ evolution neither for any powder nor for any single crystal. This indicates that the intrinsic photocatalytic activity is not caused by or connected to an enhanced visible light absorption.

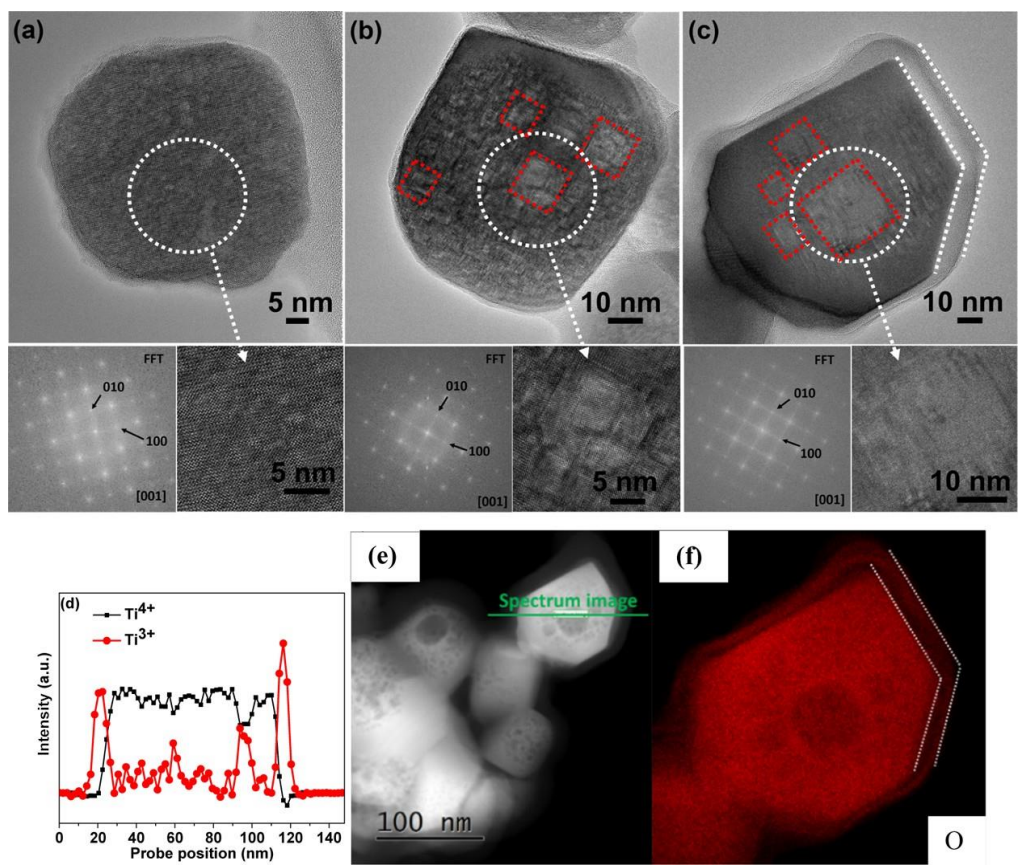

Figure 3. (a−c) TEM images for STO nanopowders (a) before treatment, (b) after annealed at 700 °C for 4 h, and (c) after annealed at 1100 °C for 4 h, respectively. (d) Distribution of $Ti^{3+}$ and $Ti^{4+}$ species across the particle, as shown in the STEM image in (e). The results are obtained from the EELS line scan data. (f) STEM−EDX mapping for oxygen.

High-resolution transmission electron microscopy (HRTEM) investigations of the differently treated powders show the formation of defects by annealing (Figures 3b,c and S5a−c). Compared with the untreated powders (Figure 3a), annealed samples show degraded areas that are formed over larger parts of the crystallites, as illustrated by the red dotted lines in the HRTEM images (Figure 3b,c). These planar-like defects give rise to diffuse streaks in the diffractograms (fast Fourier transforms) of the HRTEM images (Figure 3b,c). These lattice defects can be ascribed to the formation of oxygen vacancies and their coalescence.[45,46] Clearly, the density of these defects in the particle core increases with treatment temperature.

In addition, for particles annealed at ≥700 °C, a thin amorphous layer (≈2 nm) at the outer rim of the particle becomes visible (Figure 3b) - this layer cannot be observed for pristine STO particles (Figure 3a). The thickness of the amorphous rim increases significantly at higher annealing temperatures. For samples annealed at 1100 °C, the outer rim reaches a thickness of ≈10 nm (Figure 3c). To analyze the difference between particle bulk and rim, we performed electron-energy loss spectroscopy (EELS) and scanning TEM−energy-dispersive X-ray (STEM−EDX). An EELS line scan is shown in Figure 3d−f, and point measurements across the particle (particle in Figure 3c) are shown in Figure S5d−f. The line scan shows a strong presence of $Ti^{3+}$ in the amorphous/distorted rim around the particle core.[47] Additional STEM−EDX (Figures 3f and S5g,h) - in line with X-ray photoelectron spectrometer (XPS) data discussed below - confirms this rim to be a hydroxylated STO layer. Such a hydroxylated $Ti^{3+}$-enriched layer is absent in Ar-treated samples (Figure S6). All of the powders under all hydro- genation conditions as well as under argon treatments up to 1100 °C show no change in the crystal structure using macroscopic X-ray diffraction (XRD). The powders maintain the characteristic STO patterns (Figure 4a), but with increasing temperatures (from 500 to 1100 °C), a slight decrease of full width at half maximum (fwhm) of all reflexes can be seen (Figure S7) - this can be ascribed to the increase of the particle size because of the sintering effects (in accordance with SEM and BET characterization).

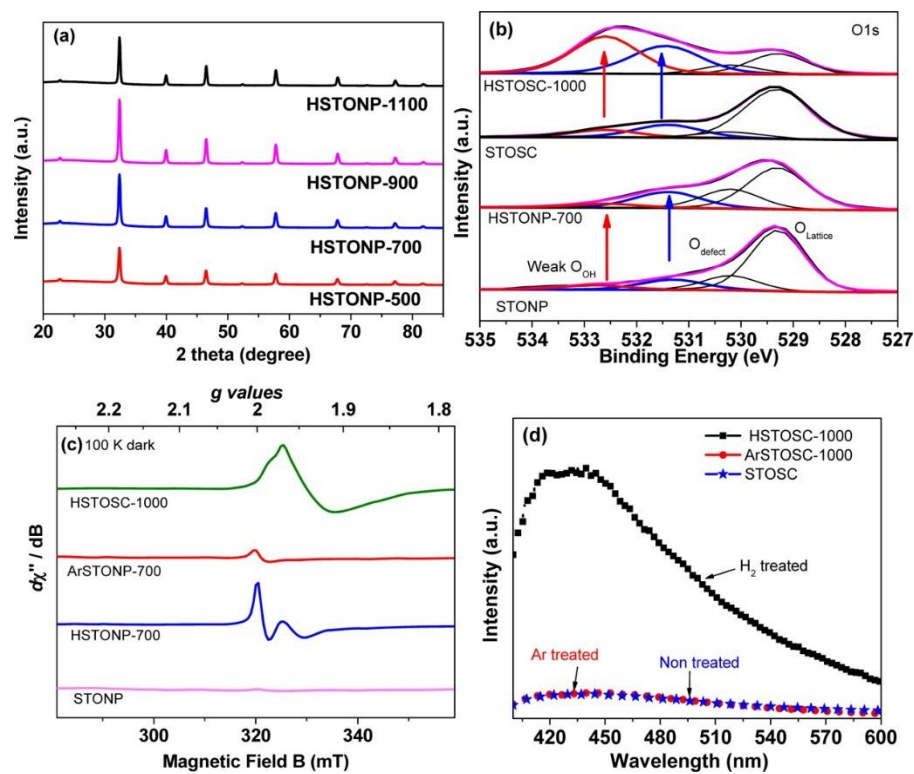

Figure 4. (a) Powder XRD patterns for representative samples obtained under different annealing conditions. (b) High-resolution XPS O 1s spectra from STOSCs before and after annealing at 1000 °C for 4 h, and STO nanopowders before and after annealing at 700 °C for 4 h. (c) EPR spectra for STOSCs and nanopowders after annealing under different conditions. (d) PL spectra for representative samples with 325 nm laser excitation in air.

In line with the above TEM observations, XPS shows that $H_2$ annealing at these temperatures causes a clear change of the surface composition of STO. Most pronounced is the formation of surface hydroxyl groups as illustrated by the changes in the O 1s peak in Figure 4b (details to the XPS characterization are given in the Supporting Information, Figures S8−S10). Generally, on STO, the O 1s peak can be deconvoluted to contributions from lattice oxide at 529.3 eV and hydroxyl species at 531.4 and 532.6 eV (Table S2).[40,48−50] The examples in Figure 4b show that for STO powder annealing in $H_2$ at 700 °C for 4 h, the amount of surface hydroxyl increases up to 28%. This alteration of the surface hydroxylation can also be clearly seen on the two types of single-crystal surfaces from the O 1s peak before and after hydrogen treatment (Figures 4b, S9, and S10). For the STO[100] single crystal, after hydrogenation, a strong increase up to 32.2% and 40.1% can be observed for the peaks at 531.4 and 532.6 eV, respectively (Figure 4b). The two hydroxide species can be ascribed to the surface-terminating hydroxyls that are bound to different surface sites;[40,48−50] including for example, hydroxyl species bound to different valence states of titanium ($Ti^{2+}$, $Ti^{3+}$).[51] As surface hydroxylation increases, accordingly a strong decrease of the lattice oxygen peak (529.3 eV) is observed.[48] A summary of the XPS-detected speciation and amounts of the different oxygen species is given in Table S2.

However, as expected, surface hydroxylation cannot be observed for samples annealed in argon the XPS O 1s peak remains at the position of the untreated single-crystal STO[100] (Figure S10). This indicates that even when the STO[100] crystal is exposed after the Ar-treatment to ambient atmosphere, no significant hydroxylation of STO takes place. We observe, however, a slight increase of a peak at 530.2 eV (from 8.0 to 9.8%) that in literature has been ascribed to surface-defective oxygen.[52,53]

To further characterize the electronic nature of the crystal defects induced by the hydrogenation and to compare them to the Ar treatment, we carried out EPR measurements for differently annealed powder samples (Figure 4c). Some data were also acquired for the STO[100] single crystals (Figure S11) for the EPR measurements, the crystals were milled to powders.

Initial EPR measurements of pristine STO powder (Figure 4c) as well as untreated, milled single crystals did not show any signal (Figure S11c), which is in agreement with the literature.[54] Annealing the powdered sample at 500 °C in a hydrogen atmosphere leads to a sole signal at a $g$ value of $g_{iso} = 2.00$, which can be ascribed to a $Ti^{3+}/O_v$ defect with a $Ti^{3+}$ in a lattice position.[33,40] However, for the treatment of STO powder in $H_2$ at $T = 700$ °C, that is, the photocatalytically most active sample, a second broad signal with $g$ values at $g \approx 1.97$ appears (Figure S11b). This second signal can be ascribed to $Ti^{3+}$ in an amorphous or distorted lattice environment.[33,40] Increasing the temperature to 900 or 1100 °C shows that the intensity of the second species increases (Figure S11a) (in line with the TEM and EELS observation of a thicker hydroxylated rim containing $Ti^{3+}$ species). When the treatment is not performed under $H_2$ but instead under argon, only species 1 can be observed with an increasing intensity for higher temperatures (Figures 4c and S11c). Similar to the powder, also for the single-crystal sample (after $H_2$ treatment and milling), a broad signature at $g \approx 1.94$ is observed for the most active single-crystal sample (Figures 4c and S11c).[55−59] The combined results from TEM and EPR thus strongly suggest lattice defects in the core of the sample (oxygen vacancies) to be responsible for the regular EPR signal (species 1), whereas $Ti^{3+}$ in the hydroxylated rim (species 2) is responsible for the additional signal corresponding to activated samples.

To gain further information on the nature of the defects induced by the $H_2$ treatment, we carried out photoluminescence (PL) measurements. To avoid common artifacts from powdered samples, we investigated STO[100] crystal surfaces before and after annealing in $H_2$ (HSTOSC-1000) and Ar (ArSTOSC-1000) at 1000 °C for 4 h (Figure 4d). The hydrogenated sample shows a strong emission band with a maximum at 430 nm, whereas neither stoichiometric STO[100] (in line with literature[60,61]) nor the Ar-treated sample shows any significant visible PL at RT. The occurrence of visible PL from STO has particularly been observed for defective STO and has been ascribed to a superposition of emission from trapped excitons and various defect-related emission bands.[62,63] This PL from the hydrogenated crystal is surface-related because it is strongly dependent on the ambient atmosphere (Figure S12).[64,65] In our case, considering the onset and maximum of the PL, the involved defect states show an emission peak at an energy of ~2.88 eV - these states thus lie around

0.3 eV below the conduction band of STO.

In this context, it is interesting to note that theoretical work using an LDA + U approach[66] shows that surface oxygen vacancies or related $Ti^{3+}$ centers in STO can induce localized electronic levels in the band gap ranging in energy from 0.3 to 1.14 eV below the conduction band minimum. Defects sufficiently close to the conduction band have been pinpointed to act as a mediator for the electron transfer to the environment.[28,67]

A main effect of the introduction of $Ti^{3+}/O_v$ defects into an STO surface is also the change in surface conductivity. Such changes in surface conductivity have been considered as a crucial point for the performance of photocatalysts because an increased surface conductivity may facilitate charge coupling between anodic and cathodic sites on a particle.[68] We investigated the effect of hydrogenation on the conductivity of STO[100] single-crystal surfaces using a four-point probe solid-state measurements (Figure 5a). Whereas for the intrinsic STO[100] single crystal, the surface resistance is too high to be reliably measured; all of the investigated hydrogenation treatments lead to a strongly increased conductivity. For samples treated at ≥500 °C in $H_2$, the resistance is reduced by several decades and drops from $R > 10^8$ Ω □ to a value of $R = 0.232$ Ω □. The Nb-doped STOSC shows already in the untreated sample a high conductivity in a comparable range ($R = 0.121$ Ω □). Hydrogenation of the Nb-doped crystal only slightly changes the conductivity ($R = 0.087$ Ω □). Hence, conductivity does not significantly affect photocatalytic $H_2$ evolution the Nb-doped material shows a comparable photocatalytic $H_2$ evolution activity to the non-Nb-doped crystal.

The results up to this point demonstrate the key effects of the hydrogenation treatment on the solid state and surface properties of STO under ex situ conditions. However, immersed in an electrolyte, the energetic situation may become dominated by the junction with the liquid phase. To obtain in situ information, we carried out photoelectrochemical $I-V$ and Mott−Schottky type of impedance measurements. These allow to determine the flat band potential of the material that reflects the position of the conduction band edge and its position relative to the $H_2O/H_2$ redox potential for the differently treated samples. In these experiments, the use of Nb-doped STO material is of utmost relevance - as the Nb-doped crystals provide a sufficient conductivity to perform electrochemistry directly on a highly defined STO[100] surface without $H_2$ treatment (or any other sort of reductive pretreatment commonly used to fabricate conductive single- crystal electrodes).[69]

Figure 5b,c shows photoelectrochemical measurements carried out with a set of single-crystal samples; that is, hydrogenated Nb-doped STO[100], untreated Nb-doped STO[100], and hydrogenated STO[100] (HNbSTOSC- 1000, NbSTOSC and HSTOSC-1000). Please note that for the nondoped STO, only the hydrogenated materials show sufficient conductivity to obtain reliable electrochemical data. For all of these samples, well-defined photocurrent spectra can be obtained with an onset of the photocurrent at wavelengths below 380 nm. An evaluation for an indirect band gap (Figure 5b inset) yields a value of 3.20 eV for all samples. This shows that hydrogenation does not lead to a change in the photoelectrochemical band gap this is in spite of the observed changes in light absorption spectra and the observed defect-related visible light PL (i.e., the generated electronic states that lead to visible light absorption in reduced STO do not contribute significantly to a photocurrent likely this can be ascribed to a lack of mobility of these visible light-generated charge carriers).

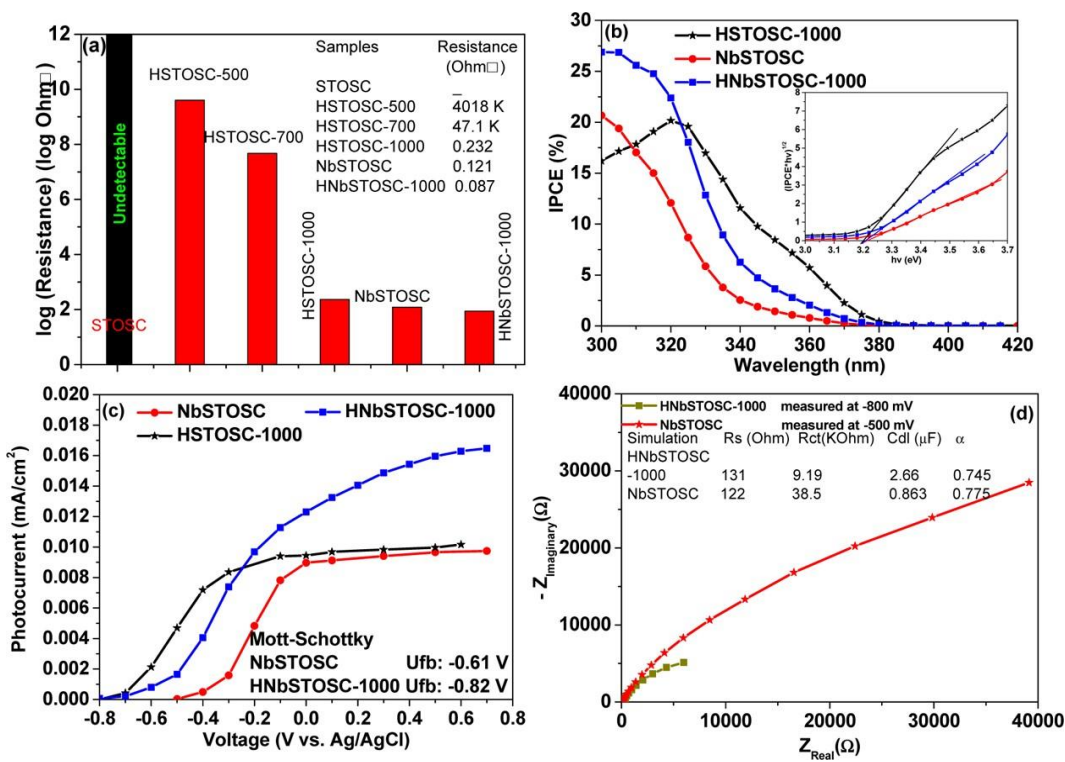

Figure 5. (a) Resistance of different crystals obtained by a four-point probe measurement (inset: experimental values). (b,c) Photoelectrochemical characterization for STOSC after annealing in hydrogen at 1000 °C for 4 h and Nb-doped STOSC before and after treatment in hydrogen at 1000 °C for 4 h. (b) Photocurrent spectra measured at +700 mV (vs Ag/AgCl), inset: band gap evaluation from photocurrent spectra in (b). (c) Photocurrent−voltage plots measured at a wavelength of 310 nm. (d) Impedance measurements for Nb-doped STOSC before and after treatment in hydrogen at 1000 °C for 4 h at their respective flat-band potentials. Inset: Simulation results for the impedance data with a Randle's equivalent circuit with a constant phase element.

Significant effects of the hydrogenation can, however, be observed in photocurrent response versus voltage measure- ments (Figure 5c). Clearly the optical flat band potential ($U_{fb}$) for the hydrogenated single crystals shows a significant shift to negative values compared to the nonhydrogenated samples ($U_{fb\text{-NbSTOSC}} = -0.50$ V, $U_{fb\text{-HSTOSC-1000}} = -0.80$ V; $U_{fb\text{-HNbSTOSC-1000}} = -0.80$ V). Please note that the $U_{fb}$ for hydrogenated STO [100] coincides with hydrogenated Nb:STO[100]. A similar shift in the flat-band potential is observed from Mott−Schottky measurements for Nb-doped STOSC samples before and after the hydrogenation treatment (Figure S13) with $U_{fb} = -0.61$ V for the nonhydrogenated Nb-doped STO and $U_{fb} = -0.82$ V for the hydrogenated material. This shift in $U_{fb}$ could, in principle, be ascribed to two effects: (i) a shift in the Fermi level because of increased doping or (ii) a shift of the surface band edge pinning of conduction and valence band ($E^0_{cb}$, $E^0_{vb}$). The fact that the nonhydrogenated but heavily Nb-doped sample (1.4 at. % Nb) does not show a shift in $U_{fb}$ indicates that doping is not the primary cause for the shift of $U_{fb}$. The shift in $U_{fb}$ is rather because of a shift of the surface band edge pinning caused by hydrogenation that induces differences in surface termination by hydroxylation.[70]

Theoretical and experimental work shows that specific defects can strongly affect the amount and type of hydroxylation of oxide surfaces[70–75] and in turn can affect $U_{fb}$ by a pinning of surface band edge positions, as illustrated in Figure S15. The $Ti^{3+}$ states embedded in the hydroxylated shell may serve as a charge-transfer mediator across this hydroxylated layer (as discussed below).[76] In terms of thermodynamics, the observed shift in $U_{fb}$ of 200 mV after hydroxylation represents a significant increase in driving force for the electron transfer from the hydrogenated STO substrates to the electrolyte and thus provides an energetic reason for the enhanced hydrogen generation.[28]

To evaluate the additional kinetic effects, we measured the charge-transfer resistance of a hydrogenated and nonhydrogenated Nb:STO crystals at their respective flat band potentials (i.e., at −800 mV versus Ag/AgCl for hydrogenated NbSTOSC (HNbSTOSC-1000) and at −500 mV versus Ag/AgCl for untreated NbSTOSC) under identical illumination conditions (Figure 5d). Evidently, the charge-transfer resist- ance for the hydrogenated sample at flat-band conditions is significantly lower. This indicates that not only thermodynamics but also kinetic effects (charge-transfer mediation) contribute to the observed strongly enhanced photocatalytic $H_2$ generation. This is in line with the results obtained from the intensity-modulated photovoltage spectroscopy (IMVS). Fig- ure S14 shows an evaluation of the recombination ($t_r$) time constants in dependence of the photon flux for the Nb:STO clearly the hydrogenated sample provides a much slower recombination kinetics than the nontreated sample. A most straightforward explanation for this kinetic effect is the mediation of the electron transfer by the $Ti^{3+}_{\square}$ states in the outer shell of the particles. Note that $Ti^{3+}/O_v$ defects created in Ar (leading to $Ti^{3+}$ in lattice positions) do not provide activation but are detrimental to photoinduced reactions (these defects therefore seem to mainly act as recombination centers). In other words, defects in the particle core (on lattice positions) are detrimental to photocatalytic $H_2$ generation $Ti^{3+}$ states in the particle rim (distorted lattice or amorphous material) are beneficial for charge transfer.

## CONCLUSIONS

The results of this study indicate that a main effect of hydrogenation of STO (in view of the photocatalytic $H_2$ generation) is the formation of a hydroxylated particle shell that leads to a beneficial shift of the band edges relative to the $H_2$ evolution potential, whereas $Ti^{3+}$ states present in the shell act as mediators for the photoelectron transfer to the aqueous environment. That is, the synergistic interplay of the hydroxylation and the formation of suitable defects leads to an intrinsic stable enhancement of STO for photocatalytic $H_2$ generation. An optimum in hydroxylation conditions exists, as for too high-treatment temperature sintering of the particles takes place that drastically reduces the active area. Other factors such as the drastic improvement of the conductivity caused by the hydrogen treatment do not play a crucial role for improving the photocatalytic activity of STO.

## SUMMARY AND PERSPECTIVE

Overall, the present work clearly shows that an optimized high- temperature $H_2$ treatment of STO nanopowders and single crystals allows for noble-metal-free and external-bias-free photocatalytic hydrogen generation from neutral methanol aqueous solutions. We find that the combination of thermodynamic effects (hydroxylation) leading to a shift of the flat band potential and kinetic effects (suitable $Ti^{3+}$ charge- transfer mediator states) acts synergistically to obtain hydrogen evolution from STO without any additional cocatalyst. Furthermore, under optimized conditions, no matter if powder samples or single crystals, the amount of hydrogen produced in this work is significantly higher (more than 50 times) than observed in the early work of Wagner and Somorjai.[26,27] The results are also comparable with recent work on mesocrystal superstructure STO.[10] However, compared to the work of Somorjai et al., the experimental conditions used here are much more moderate and appear to be of an attractive economic benefit compared to noble metal cocatalysts.

To compare the beneficial effect of hydrogenation with conventional Pt cocatalysts, we deposited 1 nm of Pt on an STOSC (note that the Pt aggregates to distinct particles, as shown in Figure S1d). For the platinized surface, we measured a $H_2$ evolution rate of 8.6 μL/(h cm$^2$) which is 14 times higher than observed for the hydrogenated surface (Figure S1c). In other words, Pt cocatalyst is still considerably more active than intrinsic activation. We note also that a combined hydro- genation and platinization is of additional benefit (Figure S1c). Compared with the reported work using Pt decoration, the present work shows a lower quantum yield;[33,40,53,77] never- theless, the presented intrinsic activation is certainly from an economic (see Table S1) and scientific view of a high interest and a number of open factors such as the geometry of the substrate (e.g., 1D-structures) and combined dopant−defect engineering strategies are possible and promising pathways to further increase this noble-metal-free approach.

## MATERIALS AND METHODS

**Materials.** STO powders (99.95% purity, 4.81 g/cm$^3$, mp 2080 °C) were purchased from the US Research Nanomaterials, Inc. Single crystals of STO [100, $K < 0.5°$, 5 mm × 5 mm ×0.5 mm, one side polished, $R_a <$ 0.5 nm, edges along (010)/(001), transparent color] and Nb-doped STOSCs [1.4 at. % Nb, (100), $K < 0.5°$, 5 mm × 5 mm × 0.5 mm, one side polished, $R_a < 0.5$ nm] were purchased from Crystal GmbH, Germany.

**High-Temperature Hydrogenation Treatment.** Two different temperature profiles (condition A and B) were used for hydrogenation (as illustrated in Figure S1e).

Condition A (temperature ramping): 100 mg of STO powders or a single crystal was placed into a pure $Al_2O_3$ boat and placed into a tube furnace. The temperature was increased from RT to a target temperature at a ramping rate of 5 °C/min. Hydrogen gas (Linde, 99.99%) was continuously flowing across the boat during the entire annealing process at a rate of 6 L/h.

Condition B (temperature stepping): 100 mg of STO powders or a single crystal were placed into a pure $Al_2O_3$ boat and placed into the tube of a movable furnace. The furnace was then preheated and after the temperature reached the target temperature, we moved the furnace over the sample zone; after annealing, the furnace was moved away. During the experiment, hydrogen gas (Linde, 99.99%) was continuously flowing during annealing at a flowing rate of 6 L/h across the sample. Alternatively, for reference experiments, argon gas (Linde, 99.99%) was used. Some reference experiments were carried out with a boat made of a yttria-stabilized zirconia (Al free); it provides the same photocatalytic result as obtained using the $Al_2O_3$ boat (see Figure S16). The single crystals (as described in the manuscript) were denoted as X(Nb)STOSC-*T*, and the nanoparticles were denoted as XSTONP-*T*: where X is the gas atmosphere, H indicates hydrogen, Ar indicates argon, and *T* is the annealing temperature.

**Characterization.**

Morphological characterization of the STO-based powders was performed using a field-emission scanning electron microscope (Hitachi SEM FE 4800). Samples were glued onto the SEM stage with a silver paste. For crystallographic characterization, XRD analysis was performed with an X'Pert Philips MPD equipped with a Panalytical X celerator detector using graphite monochromized Cu Kα radiation (α = 1.54056 Å). Information on the chemical composition of the powders was obtained using an XPS (PHI 5600, USA). XPS spectra were acquired using an Al standard X-ray source with a pass energy of 23.5 eV. All XPS element peaks were shifted to the Ti 2p position at 458.0 eV. No neutralization for the spectra was used. The analyzer resolution was set to 0.2 eV. XPS fitting was performed using a software (XPSpeak41); the fwhm was fixed between 1.2 and 1.5 eV, the Lorentzian/Gaussian ratio of peaks maintained at 20%. The ratio of different species after simulation is obtained from area ratios of deconvoluted peaks.

For PL measurements, the powder samples were pressed to pellets and excited with a 325 nm He−Cd laser. The PL spectra were recorded at RT with an iHR320 monochromator and Synergy Si CCD camera (both Horiba Jobin-Yvon). The spectra are corrected for the spectral sensitivity of the setup, determined with the help of a calibrated halogen lamp.

HRTEM was performed with a Philips CM 300 UT, Philips CM 30, and FEI Titan Themis[3] 300 equipped with image and probe correctors at an acceleration voltage of 300 kV. The HRTEM images were recorded by using a 4k complementary metal-oxide-semi- conductor camera (Ceta 16M). EELS and EDX element mapping were performed with a Gatan imaging filter (GIF Quantum) in a STEM mode at 200 kV with an energy resolution of approximately 1.0 eV. Samples were obtained by dispersion in ethanol, and sonication and drop-casting onto TEM copper grids coated with a lacey carbon film.

EPR spectra were recorded using a JEOL continuous wave spectrometer JES-FA200 equipped with an X-band Gunn oscillator bridge, a cylindrical mode cavity, and a helium cryostat. The samples were measured in air atmosphere in quartz EPR tubes with a comparable loading of ≈20 mg. The spectra shown were measured with the following parameters: temperature 100 K, microwave frequency 8.966 GHz, modulation width 1.0 mT, microwave power 1 mW, modulation frequency 100 kHz, and a time constant of 0.1 s. Analysis and simulation of the data were performed using the software "eview" and "esim" written by E. Bill (E-mail: ebill@gwdg.de, MPI for Chemical Energy Conversion, Mülheim an der Ruhr). Single crystals were milled into powders before measurements.

Four-point conductivity measurements were performed with four tungsten probes using a Probe Head Jandel 12-0401 connected with a Keithley (2401 SourceMeter). The distance between the probes was 0.115 cm. The specific resistance was calculated based on the equation

$$R_s = \frac{\pi}{\ln(2)} \frac{U}{I}$$

where $R_s$ is the specific resistance (resistance of a square sample), $U$ is the applied voltage, and $I$ is the current detected.

Reflectivity of powder samples and transmittance of single-crystal substrates are measured on a PerkinElmer LAMBDA 950 UV/Vis spectrophotometer with an integrating sphere. $Ba_2SO_4$ substrate was used as a standard reference.

Nitrogen adsorption−desorption measurements at liquid nitrogen temperature were performed using a volumetric gas sorption analyzer Nova 4200e (Quantachrome, USA). Degassing of the samples under vacuum ($p < 10^{-2}$ Torr) was performed for 6 h at 130 °C. Data were evaluated according to the BET theory (seven equidistant points, $0.02 < p/p_0 < 0.35$), the Barrett−Joyner−Halenda theory (desorption, $p/p_0 > 0.35$), and the deBoers $Vt$ plot method. The NIST recommendation for the nitrogen cross section of 0.162 $nm^2$ was used.

**Photocatalytic $H_2$ Generation**. Photocatalytic $H_2$ evolution of the different powder samples was measured under simulated AM1.5 light provided by a solar simulator (300 W xenon with optical solar light filter). The photocatalytic experiments were performed under open-circuit conditions in an aqueous methanol solution (50 vol %) under AM 1.5 (100 mW/cm$^2$) solar simulator illumination. The light intensity incident on the cell was measured using a calibrated Si photodiode. The amount of $H_2$ produced was measured from the gas phase over the illuminated liquid in a closed reactor using Shimadzu gas chromatograph GC-2010 plus with a thermal conductivity detector.

The long-term stability of $H_2$ evolution was evaluated using a hydrogenated STOSC in 24 h cycles. After each cycle, the head space of the reactor was purged, and then illumination under 365 nm LED started again. For the stability test of hydrogenated STO powders, we used the same approach as above but illuminated under AM 1.5.

To compare the hydrogenation with Pt, we used a single-crystal substrate that carried the Pt layer of a nominal thickness of 1 nm. Pt was deposited using a plasma sputter device (EMSCD500, Leica) operated at 16 mA and a pressure of $10^{-2}$ mbar. Ar was used as the moderating gas. Under these conditions, Pt decorates the STO surface as particles (Figure S1d).

To prepare suspensions for $H_2$ measurements, 2 mg of sample powders was dispersed in 10 mL of deionized water/methanol (50/50 v %) and ultrasonicated for 15 min. During illumination, the suspensions were continuously stirred. The crystals were immersed into the same electrolyte and irradiated with the respective light source. The photocatalytic $H_2$ evolution of single crystals was measured, in addition to AM 1.5 conditions, under 365 nm LED illumination (160 mW/cm$^2$, UV-LED smart SN1251, BJ2017) or using a cw-HeCd laser (325 nm, 60 mW/cm$^2$, Kimmon, Japan).

**Electrochemical Measurements**.

Electrochemical Mott−Schottky-type measurements were performed using a Zahner IM6 (ZAHNER Elektrik, Kronach, Germany) work station in the dark in 0.1 M $Na_2SO_4$. Platinum grid was used as a counter electrode, and a Ag/AgCl (3 M KCl) electrode in a Haber−Luggin capillary was used as a reference electrode. The tested potential range was −1.2 to 1.2 V. Capacitance data for Mott−Schottky plots were evaluated at 1000 Hz. To establish an Ohmic contact to the single crystal, the backside was exposed to sputtering with an Ar beam for approx. 100 nm. Then, the backside of the crystal was coated with Ga/In and connected with a Cu wire using silver glue. Impedance measurements for Nb-doped STO (1.4 at. %) single crystals before and after the hydrogenation treatments were performed in 0.1 M $Na_2SO_4$ aqueous electrolyte (15% methanol, $N_2$ saturated) with a platinum grid as a counter electrode. Measurements were performed at their respective flatband potentials, for example, at −800 mV (vs Ag/AgCl) for HNbSTOSC and at −500 mV (vs Ag/AgCl) for NbSTOSC. Impedance data were fitted using a Randle's circuit (with a constant phase element CPE: $R_\Omega/R^{CPE}$) on a Zahner IM6 workstation.

**Photoelectrochemical Measurements**.

IMVS measurements were carried out using modulated light from a high power LED ($\lambda$ = 369 nm, 90 mW). The modulation frequency was controlled by a frequency response analyzer (FRA), and the photovoltage of the cell was measured using a Zahner IM6 electrochemical interface and fed back into the FRA for analysis. The electrolyte was 0.1 M $Na_2SO_4$ with 30 vol % methanol.

Photocurrent measurements were carried out with an electro- chemical setup consisting of three electrodes:

an Ag/AgCl reference electrode, a Pt plate as a counter electrode, and the investigated sample as working electrode back contacted with a Cu wire attached by silver glue. The back contacted 5 × 5 mm$^2$ sized sample was then exposed to the electrolyte. The sample was irradiated with a 150 W Xe arc lamp (LOT-Oriel Instruments) monochromated with a Cornerstone motorized 1/8 m monochromator. The monochromatic light was focused to a spot of 5 × 3 mm$^2$ on the sample surface through electrolyte and through a quartz pass window in the electrochemical cell.

Photocurrent spectra were acquired in aqueous 0.1 M $Na_2SO_4$ at a potential of 700 mV (vs Ag/AgCl). At each wavelength, a photocurrent transient was recorded and the steady-state photo- current was assessed. Such photocurrent transients were recorded for 20 s using an electronic shutter system and A/D data acquisition. The photocurrent transients were acquired in 5 nm intervals. For photocurrent−voltage curves, the wavelength of light was fixed at 310 nm, and the voltage was changed from positive to negative.

## ACKNOWLEDGMENTS

The authors would like to acknowledge ERC, DFG, and the Erlangen DFG cluster of excellence (EAM) as well as EXC315 (Bridge) for their financial support. The authors would like to thank Dr. Ing. Gebhard Matt for the four-point conductivity measurements. The authors would like to thank Dr. Stefanie Rechberger for valuable discussions and comments on EELS results. The authors would also like to thank Dr. Stefan Romeis for the BET surface area measurements.